\documentclass[prd,twocolumn,nofootinbib,superscriptaddress]{revtex4-1}
\usepackage{preamble}

\graphicspath{{../figures/pdf-files/}}

\begin{document}

\title{Probing Massive Scalar/Vector Fields with Binary Pulsars}

\author{Brian C.~Seymour}
\affiliation{Department of Physics, University of Virginia, Charlottesville, Virginia 22904, USA}
\affiliation{Department of Applied Mathematics and Theoretical Physics, University of Cambridge, Cambridge CB3 0WA, United Kingdom}

\author{Kent Yagi}
\affiliation{Department of Physics, University of Virginia, Charlottesville, Virginia 22904, USA}

\date{\today}

\begin{abstract} 

Precision tests of general relativity can be conducted by observing binary pulsars. 
Theories with massive fields exist to explain a variety of phenomena from dark energy to the strong CP problem.
Existing pulsar binaries,  such as the white dwarf-pulsar binary J1738+0333, have been used to place stringent bounds on the scalar dipole emission, and radio telescopes may detect a pulsar orbiting a black hole in the future.
In this paper, we study the ability of pulsar binaries to probe theories involving massive scalar and vector fields through the measurement of the orbital decay rate.
With a generic framework, we describe corrections to orbital decay rate due to (a) modification of GR quadrupolar radiation and (b) dipolar radiation of a massive field.
We then consider three concrete examples: (i) massive Brans-Dicke theory, (ii) general relativity with axions, and (iii) general relativity with bound dark matter and a dark force.
Finally, we apply direct observations of J1738 and simulations of a black hole-pulsar binary to bound theory parameters such as field's mass and coupling constant.
We find new constraints on bound dark matter interactions with PSR J1738, and a black hole-pulsar discovery would likely improve these further. Such bounds are complementary to future gravitational-wave bounds. Regarding other theories, we find similar constraints to previous pulsar measurements for massive Brans-Dicke theory and axions.
These results show that new pulsar binaries will continue to allow for more stringent tests of gravity.

\end{abstract}
\maketitle

\section{Introduction}

At astrophysical scales, gravity is the sole long-range force, and it is best described by general relativity (GR) \cite{Will:2014kxa,Berti:2015itd}. However, unobserved long-range interactions could be a part of the unknown physics--dark matter, dark energy, beyond Standard Model. One such long-range modification could be from (light) massive scalar and vector fields in our Universe. An additional massive field could change the physics at astrophysical and cosmological scales differently because the field is screened out above a characteristic scale corresponding to the Compton length of the field's mass.

Astrophysical observations can find or constrain the existence of a massive scalar/vector field by verifying the consistency of GR as the sole force. So far, GR has been strongly constrained by solar system tests \cite{Will:2014kxa}, and cosmological observations are ongoing to study the gravity at large scales \cite{Koyama:2015vza}. On the other hand, precision measurements of binary pulsars (PSRs) can be made through PSR timing to test the strong field regime of gravity \cite{Wex:2014nva,Stairs:2003eg}. Furthermore, gravitational wave (GW) observations can probe the strong field effects of gravity in a dynamical setting \cite{TheLIGOScientific:2016src,Abbott:2018lct,Yunes:2016jcc,LIGOScientific:2019fpa,Berti:2018cxi,Berti:2018vdi,Nair:2019iur,Yamada:2019zrb,Tahura:2019dgr}. While we know GR is an effective field theory (EFT) and must break down at a certain energy \cite{Will:2014kxa,Berti:2015itd}, extra degrees of freedom may become relevant at higher energy scales in a gravitational theory. Thus, testing GR is a way to look for signatures of new physics.

Let us now focus on binary PSR tests of gravity \cite{Stairs:2003eg,Wex:2014nva}. These systems consist of a PSR with a neutron star (NS), white dwarf (WD), or even another PSR. Therefore, measurements of PSR binaries allow us to study a system of strongly self-gravitating objects. On the other hand, the binary is widely-separated, and correspondingly the velocities of the bodies are slow compared to the speed of light (typically by a factor of $10^{-3}$ or smaller). Compared to GW sources, PSR binaries are less dynamical which allows their evolution to be more analytically tractable. Tests of GR with binary PSRs are carried out with the measurement of parameterized post-Keplarian (PPK) parameters. Radio astronomy measurements of two PPK parameters give the binary masses, and any additional PPK measurement can be used to test GR. 

Previously, we have studied tests of GR with PSR observations with a BH-PSR binary and a triple PSR system. First, we found the discovery of a BH-PSR binary allows stringent tests for some modified theories of gravity \cite{Seymour:2018bce}. We studied both the orbital decay rate measurement of a stellar mass BH-PSR binary and the quadrupole measurement of Sgr A* by a closely orbiting PSR. We saw that a BH-PSR binary is strongest (compared to NS-PSR or WD-PSR) for corrections entering at lowest post-Newtonian (PN) orders\footnote{A PN order of $n$ refers to a correction proportional to $\left(v / c \right)^{2n} $ relative to the leading GR term. }. The detection of a BH-PSR binary would most stringently test a varying gravitational constant (at $-4$PN) \cite{Seymour:2018bce} and theories with scalar degrees of freedom allowing scalar dipole radiation (at $-1$PN), such as Einstein-dilaton Gauss-Bonnet gravity \cite{Yagi:2015oca}. Second, we probed massive scalar fields with strong equivalence principle (SEP) violation constraints of the hierarchical triple PSR system, PSR J0337+1715 \cite{Seymour:2019tir}. Additional scalar degrees of freedom induce SEP violation, and we used this to constrain some massive scalars. We were able to place the strongest constraint on massive Brans-Dicke theory to date. Thus, we continue our previous work here and investigate possible binary PSR constraints with a BH-PSR or WD-PSR. 

In this paper, we will examine tests of massive scalar/vector fields primarily with two such binaries. The first binary system is PSR-WD J1738+0333 (henceforth PSR J1738) \cite{Freire:2012mg}. It is currently one of the best binary PSR systems known to date to constrain dipolar radiation. It has a precise measurement of the intrinsic orbital decay rate which is consistent with the GR prediction. These features make it a prime test of massive fields.

The second system considered in this paper is a BH-PSR with the Square Kilometer Array (SKA) \cite{Liu:2014uka}.
A BH-PSR binary has not been found with radio astronomy. However, SKA is a next-generation radio telescope which will vastly increase the odds of finding one. Previous work on population analysis has estimated there are around 3-80 BH-PSR binaries in the Galactic disk, and an instrument like Five-hundred-meter Aperture Spherical radio Telescope could detect up to 10\% of them \cite{Shao:2018qpt}. In this paper, we will consider a BH in a binary with a millisecond PSR because it allows a much higher timing precision and a more accurate measurement of binary parameters. To make projected bounds for a BH-PSR binary, we will use the results of Ref.~\cite{Liu:2014uka}, which forecast measurement accuracies with SKA. The binaries in the simulation have a distribution of periods, but we picked a specific period of 3 days. We also use $10 \Msolar$ and $1.4 \Msolar$ for the BH and PSR, and other binary parameters are discussed in App.~\ref{app:binary-param}. In comparison to a PSR-WD binary, a BH-PSR binary has the strength of very precise orbital decay rate measurement due to the significantly higher GW radiation. 

We will begin by describing a generic formalism for the orbital decay rate in a binary with additional massive fields. This formalism will capture the leading order modification away from solely GR. After this, we will consider some example theories in the formalism and find the resulting constraints from PSR J1738 and BH-PSR with SKA. 

We will consider two specific theories with massive scalar fields: (i) massive Brans-Dicke (MBD) theory \cite{Alsing:2011er,Berti:2012bp} and (ii) axions in GR \cite{Hook:2017psm,Huang:2018pbu}. We also consider a theory with a massive vector field, namely (iii) GR with bound dark matter and a dark matter interaction \cite{Alexander:2018qzg}. MBD theory is a scalar-tensor theory of gravity and a generalization of the original Brans-Dicke (BD) theory with a massless scalar field \cite{Brans:1961sx}. The axion is a hypothetical particle that aims to solve the strong CP problem \cite{Peccei:1977hh} in QCD\footnote{The XENON1T experiment has found excess events over the background around 2-3 keV \cite{Aprile:2020tmw}. One possible explanation for this observation is the measurement of solar axions, which is at the 3.5$\sigma$ significance. }. Strongly gravitating objects can acquire significant axion charges so its presence could show up in astrophysical observations \cite{Hook:2017psm}. Finally, we examine a generic description of dark matter interactions. Dark matter could become bound inside compact objects \cite{Goldman:1989nd,Kouvaris:2010vv}, and a (massive) dark matter mediator would create a Yukawa term in the binary's energy. 

Let us now discuss our results. In MBD, we find that PSR J1738 and a BH-PSR binary can place constraints that are weaker than our previous PSR J0337 SEP violation constraints \cite{Seymour:2019tir}. We find that our binary PSR constraints on the axion are similar to previous ones of PSR J0737 \cite{Hook:2017psm} and PSR J0337 \cite{Seymour:2019tir}. Regarding the dark force in a binary, \textit{we can place constraints in a new region that characterizes dark charge of a NS}. While future GW observations can bound wavelengths $10^2 - 10^8 $ km \cite{Alexander:2018qzg}, binary PSRs can constrain wavelengths above $10^{10} $ km. We find that PSR J1738 places strong constraints on this region, and the discovery of a BH-PSR would allow an improvement of an order of magnitude. 

The rest of the paper is organized as follows. In Sec.~\ref{sec:formalism}, we describe our formalism for the modification of the orbital decay rate due to the addition of the new massive field. We discuss the calculation of the orbital decay rate modifications and how we use this to constrain theory parameters with PSR measurements. After this, we focus on each theory specifically in a different section and present bounds possible with binary PSRs: MBD in Sec.~\ref{sec:MBD}, axion in Sec.~\ref{sec:axion}, and dark force from bound dark matter in Sec.~\ref{sec:bounddm}. Finally, we conclude and discuss future work in Sec.~\ref{sec:conclusion}. We also include appendices for our GW bounds on MBD in App.~\ref{sec:MBDGW} and binary PSR parameters in App.~\ref{app:binary-param}. For the remainder of the paper, we use geometric units of $c=1$ and $G=1$.

\section{Formalism}\label{sec:formalism}

In this section, we explore an effective formalism that describes the modification of the orbital decay rate in a binary. We introduce theory dependent constants that characterize the quadrupolar and dipolar radiation. Table~\ref{tab:theorylisting} then gives the theory constants for MBD, an axion, and a dark matter mediator. We then explicitly show how a massive field modifies Kepler's third law and the orbital decay rate of the binary. Finally, we describe our method for constraining massive fields with observations of the orbital decay rate.

\subsection{Generic Formalism for Orbital Decay Rate}\label{sec:genericform}
We begin by constructing a generic formalism to describe the modification of the orbital decay rate due to the presence of an additional massive field. Recall that in GR, the gravitational radiation causes an energy loss of the binary, and in turn, this reduces its period. In GR without an additional massive field, the orbital decay rate is given by the Peters-Matthews formula 
\begin{equation}\label{eq:pdotgr-period}
    \dot P_\GR = -\frac{192 \pi}{5} \mu M^\frac{2}{3}\left( \frac{P}{2\pi}\right)^{-\frac{5}{3}} \, ,
\end{equation}
where $M$ is the total binary mass, $\mu$ is the reduced mass, and $P$ is the orbital period \cite{Peters:1963ux}. We consider a quasicircular binary in this formula because there are many binaries with negligible eccentricities (e.g.~PSR J1738 has $e=3\times 10^{-7}$). In an alternative theory of gravity, however, there can be additional radiation from scalar/vector fields and corrections to the quadrupole formula. This extra radiation will cause the orbital decay rate to differ from the standard GR case. The orbital decay rate is thus modified as 
\begin{equation}
    \dot P = \dot P_\GR \left( 1+\delta\dot P\right) \, ,
\end{equation}
where $\dot P$ is the orbital decay rate in the modified theory of gravity, and $\delta \dot P$ is the fractional deviation from the standard GR orbital decay rate. 

We will work in a PN framework to have a generic expression for $\delta \dot P$ that captures leading effects due to massive fields.
The PN formalism is a convenient way of describing deviation from GR without an additional massive field. 
The PN order describes the dependence of the relative velocity $v$ of a binary on the correction. A modification to the orbital decay rate is said to have a correction of $n$PN order relative to the leading GR term if the correction enters at $v^{2n}$. PSR binaries generally have $v\sim0.001$.

Massive fields induce correction to orbital decay rate in two main ways: dipole radiation and modification of the quadrupole radiation term. Dipole radiation enters at $-1$PN order relative to the standard GR quadrupole radiation, while the leading correction to the quadrupolar radiation enters at the relative $0$PN order. Following the typical convention used in scalar-tensor theories, the orbital decay rate in the presence of additional scalar/vector fields can be expressed as
\begin{equation}\label{eq:genericODRwKappa}
    \dot P = \dot P_\GR \left( \mathcal G^{-4/3} \frac{\kappa_1}{12}+\frac{5}{96}\frac{\kappa_D}{\mathcal G^{1/3}}\mathcal S^2 v^{-2} \right)\,,
\end{equation}
where $\kappa_1$ and $\kappa_D$ are theory-dependent constants characterizing the quadrupolar and dipolar emission respectively, while $\mathcal S$ is the difference between the sensitivities or the dimensionless scalar charges, and $\mathcal{G}$ is an effective gravitational constant containing the effect of the extra field. 

Rewriting the above equation in terms of the direct observables, one finds the leading correction to $\dot P$ as\footnote{Although the total mass in the dipole term is not a direct observable, it can be determined from other PPK parameters under the GR assumption since MBD corrections to such parameters enter at higher order than $-1$PN.}
\begin{equation} \label{eq:mbdodrconstraint}
    \delta \dot P = -1+\frac{\kappa_1}{12}\mathcal{G}^{-\frac{4}{3}} +\frac{5\kappa_D}{96 \mathcal{G}}  \left( \frac{2\pi M }{P}\right)^{-\frac{2}{3}}\mathcal{S}^2   \, ,
\end{equation}
where we used the modified Kepler's third law discussed in Sec.~\ref{sec:calculationodr}.
Although the 1PN correction to the dipole emission effect enters at the same PN order as the leading quadrupole emission one, we do not consider such an effect since it can never be the dominant correction. 
For a binary PSR with a WD or BH companion, the scalar dipole radiation correction dominates the correction to the standard GR orbital decay rate since it enters at lower PN order. On the other hand, the 0PN correction may dominate the $-1$PN dipole emission for a binary PSR with a NS companion since $\mathcal{S}^2$ nearly vanishes. By using measurements of the orbital decay rate of a binary PSR, we can bound theory parameters by comparing their deviation from observation. 

With this formalism set in place, we calculated the orbital decay rate parameters $\kappa_1$ and $\kappa_D$ explicitly for some theories involving massive scalar or vector fields. In Table~\ref{tab:theorylisting}, $\kappa_{1}$ and $\kappa_D$ are given for MBD theory, GR with axion, and GR with dark force. This formalism allows one to see the similarities of how the theories modify the orbital decay rate. Each expression in this table is derived in the section in which we fully consider the theory, but this table provides a summary of corrections to $\dot P$ in each theory.

\begingroup
\squeezetable
\begin{table*}
\begin{centering}
\begin{tabular}{c|c|c|c|c}
\hline 
\hline 
    \multirow{2}{*}{Theory}&\multirow{2}{*}{$\kappa_1$}&\multirow{2}{*}{$\kappa_{D}$}&Theoretical & Refs\\& & &  parameters &  \\ 
    \hline \hline
    \multirow{2}{*}{Massive Brans-Dicke (Sec.~\ref{sec:MBD}) }&\multirow{2}{*}{ 
     $\mathcal{G}^2 \left[ 12-6 \xi +\xi \Gamma^2\left(1 - \frac{m_s^2}{4\omega^2} \right)^2 \Theta(2\omega -m_s)  \right]$ 
     }&\multirow{2}{*}{$2 \mathcal{G} \xi \left(1 - \frac{m_s^2}{\omega^2} \right)\Theta(\omega-m_s)$ } &\multirow{2}{*}{($m_s$, $\omega_\BD$) }&\multirow{2}{*}{\cite{Berti:2012bp}}  \\ 
     & & &  & \\ 
    \multirow{2}{*}{Axion (Sec.~\ref{sec:axion})}&\multirow{2}{*}{ 
     $12 \mathcal{G}^2$ 
     }&\multirow{2}{*}{$ \frac{1}{8\pi}\mathcal{G}\left(1-\frac{m_a^2}{\omega^2}\right)^{3/2}\Theta(\omega-m_a) $} &\multirow{2}{*}{($m_a$, $f_a$) }&\multirow{2}{*}{\cite{Hook:2017psm,Huang:2018pbu}}  \\ 
     &  & & & \\ 
    \multirow{2}{*}{Dark matter mediator (Sec.~\ref{sec:bounddm})}&\multirow{2}{*}{ 
     $12 \mathcal{G}^2$ 
     }&\multirow{2}{*}{$ 2 \mathcal{G} \Theta\left(  \omega -m_v \right) $}& \multirow{2}{*}{($m_v$, $\alpha$) }&\multirow{2}{*}{\cite{Alexander:2018qzg}}  \\ 
     & & &  & \\ 
    \hline
    \hline 
    \end{tabular} 
    \end{centering}
    \caption{\label{tab:theorylisting} 
   Orbital decay rate constants $\kappa_1$ and $\kappa_D$ of Eq~\eqref{eq:genericODRwKappa} for each theory we consider in the paper. These constants correspond to the quadrupole correction and the dipole emission respectively. We also list the theoretical parameters for each theory and give references. 
    } 
\end{table*}
\endgroup

\subsection{Calculation of Orbital Decay Rate}\label{sec:calculationodr}

Next, let us examine how the orbital decay rate is modified in a binary due to an additional massive scalar/vector field. The massive field has a potential that is described by a Yukawa potential to leading order
\begin{equation}
    V_\phi \propto -m_1 m_2 \frac{q_1 q_2}{r} e^{-m_s r} \, ,
\end{equation}
where $m_i$ is the mass of body $i$, $q_i$ is the dimensionless charge of body $i$, $m_s \equiv \tilde m_s/\hbar$ is the rescaled mass of the extra field with mass $\tilde m_s$ \cite{Berti:2012bp}\footnote{For the rest of the paper, we will not distinguish between $m_s$ and $\tilde m_s$, which amounts to effectively setting $\hbar = 1$.}, and $r$ is the binary separation. In the Newtonian limit, the combined potential is described as
\begin{eqnarray}
\label{eq:V_calG}
    V = -\mathcal{G}\frac{M \mu}{r}\, , \quad
    \mathcal{G}(r) \equiv G \left( 1+\frac{V_\phi}{V_\GR}\right) \, ,
\end{eqnarray}
where $M$ and $\mu$ are the total mass and reduced mass respectively while $V_0$ is the Newtonian potential. Furthermore, in this paper we will be considering light scalar fields (i.e. $m_s r \ll 1$) so $\mathcal{G}$ is independent of $r$ and becomes an effective gravitational \emph{constant} of a binary. In this regime, we apply the virial theorem and the quasicircular approximation to see that the total energy of the binary is
\begin{equation}\label{eq:virialtheoremresults}
    E = -\frac{\mathcal{G}}{2}\frac{M \mu}{r}  = - \frac{1}{2}\mu v^2 \, .
\end{equation}
Thus, we find a modified Kepler's third law,
\begin{equation}\label{eq:modifiedkeplerlaw}
    r^3 =\frac{\mathcal{G} M}{4 \pi^2} P^2 \, ,
\end{equation}
which allows us to write orbital separation $r$ in terms of orbital period $P$. This will be crucial for changing variables containing orbital separation dependencies into those containing only period dependencies (c.f.~Eq.~\eqref{eq:mbdodrconstraint}).
Notice that within the assumption of $m_s r \ll 1$, the conservative sector (binding energy and Kepler's law) is the same as the massless case. We will see later that the dissipative sector (scalar/vector emission) acquires the field's mass dependence.

Now let us compute modifications to orbital decay rate due to additional radiation of a massive field. We compute the orbital decay rate by using the chain rule to relate it to radiation 
\begin{equation}\label{eq:PdotChainRule}
    \dot P =  \frac{d E}{d t}\left(\frac{d E}{d \omega}\right)^{-1}\frac{d P}{d \omega} \,,
\end{equation}
where $\omega = \pi f_\GW$ is the orbital angular frequency.
From Eq.~\eqref{eq:virialtheoremresults}, we compute $\frac{d E}{d \omega} = \frac{2}{3} \frac{E}{\omega}$ . Putting this into Eq.~\eqref{eq:PdotChainRule}, we find that orbital decay rate is directly related to energy loss rate,
\begin{equation}\label{eq:pdot-generic}
    \frac{\dot P}{P} = -\frac{3}{2}\frac{\dot E}{E} \, .
\end{equation}
Therefore, the existence of other fields will modify the orbital decay rate from what standard GR predicts. One final step is to define the fractional orbital decay rate modification 
\begin{equation}
    \delta \dot P\equiv \frac{\dot P - \dot P_\GR}{\dot P_\GR} = \frac{\frac{\dot E}{E}-\frac{\dot E_\GR}{E_\GR}}{\frac{\dot E_\GR}{E_\GR}} \, .
\end{equation}
Measurements of binary PSRs will place bounds on the $\delta \dot P$, which will in turn constrain theoretical parameters.

\subsection{Constraints from Orbital Decay Rate}\label{sec:ODRmethod}

Using the results from Secs.~\ref{sec:genericform} and \ref{sec:calculationodr} for the orbital decay rate in modified theories of gravity, we next find how measurements of binary PSRs can be used to constrain their theory parameters. 

PSR timing is used to measure the orbital decay rate $\dot P_\obs$. However, the measurement of the orbital decay rate can be only made to a certain statistical precision. The fractional 1-$\sigma$ statistical error in orbital decay rate will be denoted as $\delta_\STAT$. Using this, we use the following constraint with an $n$-$\sigma$ uncertainty:
\begin{equation}\label{eq:intialbounding}
    \left|\frac{\dot P_\obs - \dot P_\theo}{\dot P_\GR}  \right| < n \, \delta_\STAT \, ,
\end{equation}
where we use $\dot P_\theo$ to represent orbital decay rate with an additional massive field. 
Furthermore, an orbital decay rate measurement will slightly differ from that predicted from GR without an additional massive field. This discrepancy will be called the fractional systematic error $\delta_\SYST$,
\begin{equation}
    \dot P_\obs = \dot P_\GR(1+\delta_\SYST) \, .
\end{equation}
The fractional systematic error characterizes how well the predictions of GR without an additional field match our measurements. Using this, we can simplify Eq.~\eqref{eq:intialbounding} to be,
\begin{equation}\label{eq:mainbounding}
    \left|\delta \dot P_\theo - \delta_\SYST \right|< n 
    \delta_\STAT \, .
\end{equation}
Henceforth, we will place constraints at $95\%$ level ($n=2$). Note that for all the PSR binaries we examine, the statistical error dominates the systematic error ($\delta_\STAT \gg \delta_\SYST$)\footnote{Conversely, a higher systematic error would signify astrophysical systematics that were not accounted for, or beyond (standard) GR effects.}. In this way, we will use this in combination with the formalism for orbital decay rate modification in Eq.~\eqref{eq:mbdodrconstraint} to constrain modifications due to an additional massive field (for $\delta \dot P_\theo$ depends explicitly on theory parameters).

\section{Massive Brans-Dicke Gravity} \label{sec:MBD}

\subsection{Theory}
Let us motivate MBD theory by looking at a general action for scalar-tensor gravity\small
\begin{align}
    S &= \frac{1}{16 \pi} \int d^4x \sqrt{-g}\left[ \phi R - \frac{\omega(\phi)}{\phi}  \phi_{,\mu}\phi^{,\mu}+\tilde M(\phi)\right] +S_M(g^{\mu\nu},\phi) \, , 
\end{align}\normalsize
with scalar curvature $R$, scalar field $\phi$, coupling function $\omega(\phi)$, metric $g^{\mu \nu}$, cosmological function $\tilde M(\phi)$, and matter field action term $S_M(g^{\mu\nu},\phi)$ \cite{Clifton:2011jh}. The cosmological function adds two important effects to this theory \cite{Alsing:2011er}. First, it acts like a cosmological constant in MBD (e.g. $\Lambda(\phi) = - \tilde M(\phi)/2$). Second, it causes the scalar field to have a mass $m_s$, and the scalar field is described by a Yukawa term. Thus, this scalar gives a way to describe cosmic acceleration of the universe at large scales $r \gg m_s^{-1}$, while scales less than the characteristic length $r < m_s^{-1}$ of the scalar are modified differently.

For the rest of the article, we will consider a specific case of MBD theory. First, we restrict ourselves to a constant coupling function $\omega_\BD = \omega(\phi)$: this matches the much studied massless Brans-Dicke (BD) theory \cite{Brans:1961sx}. Furthermore, we assume that the BD field has a nonzero value $\phi_0$ determined by its cosmological evolution. Thus, the scalar field at any point can be described as $\phi = \phi_0+\varphi$, for some small perturbation $\varphi$. One can show that this formulation of MBD acquires a mass term equal to,
\begin{equation}
    \tilde M(\phi) = \frac{1}{2}\tilde M''(\phi_0)\varphi^2 + \mathcal{O}(\varphi^3)\, .
\end{equation}
The reason that we start at $\tilde M''(\phi_0)$ is because asymptotic flatness requires that $\tilde M(\phi_0)=\tilde M'(\phi_0)=0$ \cite{Alsing:2011er}. This leads to the scalar field mass equal to,
\begin{equation}
    m_s^2 =  - \frac{\phi_0}{3+2\omega_\BD} \tilde M''(\phi_0) \,,
\end{equation}
and the cosmologically imposed value of $\phi_0$ is
\begin{equation}
    \phi_0 = \frac{4 + 2 \omega_\BD}{3 + 2 \omega_\BD}\, .
\end{equation}

Lastly, the matter field contribution to the action is described in the following way. For a system of point-like masses, the matter field contribution is equal to 
\begin{equation}
    S_M(g^{\mu\nu},\phi) = - \sum_i \int d \tau_i \, m_i(\phi) \, ,
\end{equation}
where $m_i$ is the mass of each particle and $\tau_i$ is the proper time for particle $i$. The gravitational constant $G$ depends on the scalar field and can be expressed as $G = \phi_0 / \phi$. The mass of an object is influenced by the scalar field: the mass of body $i$ is equal to \cite{Alsing:2011er}
\begin{align}
    m_i(\phi) &= m_i(\phi_0)\left[1+
    s_i\left(\frac{\varphi}{\phi_0}\right)\right.\nonumber\\
    &\left.-\frac{1}{2}(s_i'-s_i^2+s_i)\left(\frac{\varphi}{\phi_0}\right)^2 +\mathcal{O}\left(\left(\frac{\varphi}{\phi_0}\right)^3\right)
    \right] \, ,
\end{align}
where the first and second sensitivities are defined to be 
\begin{align}\label{eq:sensitivities}
    s_i \equiv - \left.\frac{\partial (\ln m_i)}{\partial(\ln G)}\right\rvert_{\phi_0}\,, \quad
            s_i' \equiv - \left.\frac{\partial^2 (\ln m_i)}{\partial(\ln G)^2 }\right\rvert_{\phi_0} \,.
\end{align}
With the sensitivities defined, we can now use this information to understand the dynamics of a binary in MBD.

\subsection{Binary Pulsar Bounds }

Let us first calculate the orbital decay rate in MBD in order to place constraints with binary PSR measurements. For a binary, it is useful to define a couple of parameters that are standard in the literature in order to simplify expressions \cite{Alsing:2011er},
\begin{subequations}
    \begin{align}
        \xi &\equiv \frac{1}{2+\omega_\BD} \, ,  \\
        \Gamma &\equiv 1-2\frac{s_1 m_2+m_1s_2}{M} \, , 
    \end{align}
\end{subequations}
where $M = m_1+m_2$ is the total binary mass. If one considers the potential between two objects, the binary's effective gravitational constant is 
\begin{align}
    \mathcal{G} = 1 - \frac{1}{2}\xi &\left[ 1 - \left( 1- 2s_1 \right)\left( 1- 2s_2 \right) e^{- m_s r}\right] \, ,
\end{align}
where we then take the light scalar field limit $m_s r \ll 0 $ and find 
\begin{equation}
    \mathcal{G} \equiv 1-\xi\left(s_1+s_2-2s_1s_2\right) \, .
\end{equation}
With this gravitational constant identified, the modified Kepler's third law has the form of Eq.~\eqref{eq:modifiedkeplerlaw}. 

Next, we calculate the orbital decay rate in MBD using Eq.~\eqref{eq:pdot-generic}. To do this, we must know the amount of radiation emitted in a binary due to GW and scalar fields. Ref.~\cite{Alsing:2011er} calculated the contributions for this: GW radiation (including modification of mass quadrupole moment), scalar dipole radiation, and scalar quadrupole radiation. The expressions for these radiation sources are

\begin{subequations}\label{eq:mbdradiation}
    \begin{align}
        \dot E_Q &= -\frac{32}{5}\frac{\mathcal{G}^2 \mu^2 M^2 v^2}{r^4}\left(1-\frac{1}{2}\xi\right) \, , \\
        \dot E_{sD} &= -\frac{\mathcal{G}^2 \mu^2 M^2 \xi}{r^4}\left[\frac{2}{3}\mathcal{S}^2\left(1 - \frac{m_s^2}{\omega^2} \right)\Theta(\omega-m_s)\right] \, , \\
        \dot E_{sQ} &= -\frac{\mathcal{G}^2 \mu^2 M^2 \xi}{r^4}\left[\frac{8}{15}\Gamma^2 v^2\left(1 - \frac{m_s^2}{4\omega^2} \right)^2\Theta(2\omega-m_s)\right] \, ,
    \end{align}
\end{subequations}
where $\dot E_Q$, $\dot E_{sD}$, and $\dot E_{sQ}$ are the energy loss rate due to mass quadrupole, scalar dipole, and scalar quadrupole respectively. $\mathcal{S}^2 = (s_1-s_2)^2$ is the sensitivity difference squared, and $\Theta$ is the Heaviside step function\footnote{The step function arises in the following manner. The scalar field propagates in a region with no sources with the following field equation: $(-\Box +m_s^2)\varphi = 0$. This implies $\omega^2 = \vec{k}^2+m_s^2$. Since $\vec k^2$ is non-negative, the condition $\omega\geq m_s$ arises. Similarly, the scalar quadrupolar radiation has a condition that $2\omega\geq m_s$.}.

With the expressions for radiation in MBD, we now can calculate the orbital decay rate. We use Eq.~\eqref{eq:pdot-generic} in conjunction with the expressions for radiation in Eq.~\eqref{eq:mbdradiation}. For a quasicircular binary, the leading corrections to the orbital decay rate are given by Eq.~\eqref{eq:mbdodrconstraint} with
\allowdisplaybreaks
\begin{align}
    \kappa_1 &= \mathcal{G}^2 \left[ 12-6 \xi +\xi \Gamma^2\left(1 - \frac{m_s^2}{4\omega^2} \right)^2 \Theta(2\omega -m_s)  \right]\, , \nonumber\\
    \label{eq:MBDkappas}
    \kappa_D &= 2 \mathcal{G} \xi \left(1 - \frac{m_s^2}{\omega^2} \right)\Theta(\omega-m_s)\, .
\end{align}
With this result in hand, we can obtain constraints of MBD with binary PSR measurements.

The bounds from orbital decay rate measurement will proceed as described in Sec.~\ref{sec:ODRmethod}. There are two theory parameters ($m_s$ and $\omega_\BD$) to be bounded. An experimental observation of the orbital decay rate will therefore create constraints on the theory parameter space. This is done by using the inequality from Eq.~\eqref{eq:mainbounding}. 

\begin{figure}[htbp]
    \begin{center}
        \includegraphics[width=\linewidth]{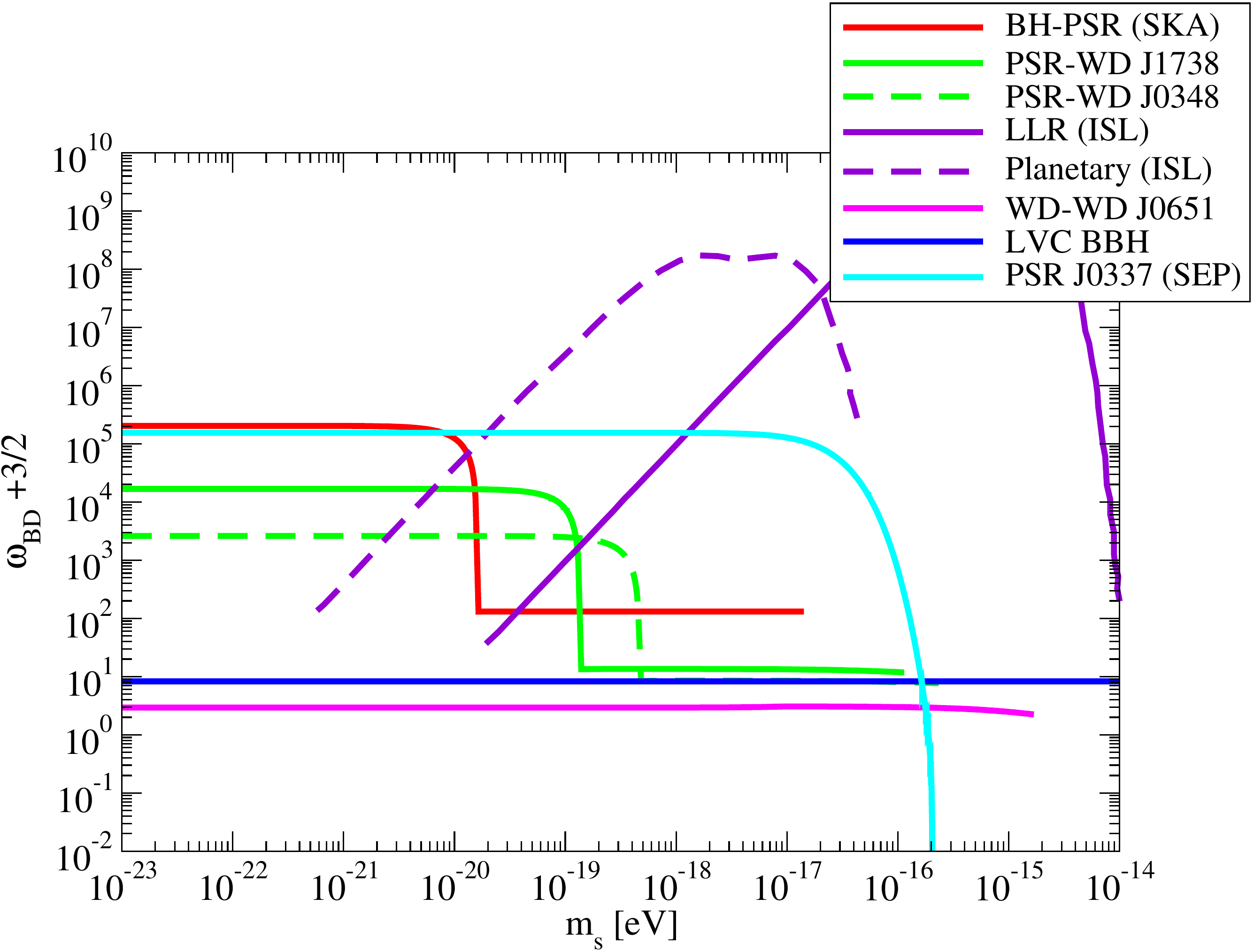}
        \caption{\label{fig:MBDOmegaBound} 
        The upper bound on the BD parameter $\omega_\BD$ as a function of the scalar field's mass $m_s$. A larger $\omega_\BD$ corresponds to a stronger bound because GR is the limit as $\omega_\BD \rightarrow \infty$.
        We use simulated constraints from a BH-PSR binary through measurements with SKA (red) \cite{Liu:2014uka}. The green lines are constraints from current PSR-WD binaries: J1738 (solid green) \cite{Freire:2012mg} and J0348 (dashed green) \cite{Antoniadis:2013pzd}. 
        We also present the bounds by the double WD (pink) \cite{Hermes:2012us} and LIGO-Virgo Catalog GWTC-1 of binary BH (blue) \cite{LIGOScientific:2019fpa}.
        We show the constraints arising from the strong equivalence principle (SEP) measurement of the PSR triple system J0337 (cyan)~\cite{Seymour:2019tir}, and due to inverse square law measurements (ISL) in purple: LLR (solid purple) \cite{Williams:2004qba} and planetary measurements (dashed purple) \cite{Adelberger:2009zz}.
        }
    \end{center}
\end{figure}

We will now use this formulation in conjunction with binary PSR measurements. 
We use measurements of PSR-WD J1738 \cite{Freire:2012mg}, PSR J0348 \cite{Antoniadis:2013pzd}, WD-WD J0651 \cite{Hermes:2012us}, and BH-PSR simulations \cite{Liu:2014uka} in conjunction with the constraints from Eq.~\eqref{eq:mbdodrconstraint}. These full binary parameters are given in Table~\ref{tab:binary-values}. We plot the results of these constraints in Fig.~\ref{fig:MBDOmegaBound}.

Let us now put our PSR bounds in context and compare it to other methods of constraining MBD. In Fig.~\ref{fig:MBDOmegaBound}, we have color-coded each bound according to the type of system. Binary bounds are shown for PSR-WD (green) and simulated BH-PSR (red), where one can see a stronger bound when $m_s < \omega$ due to scalar dipole emission but is bounded by quadrupolar modification for $m_s > \omega$. Dipolar emission for WD-WD (pink) is suppressed since the binary components are nearly identical. Bounds from PSR-WD, BH-PSR, and WD-WD halt around $m_s\sim 10^{-16} $ eV since the approximation $m_s r \ll 1$ breaks down. The best GW bounds are from LIGO-Virgo BBH Catalog GWTC-1 (blue) \cite{LIGOScientific:2019fpa}. We derive these bounds on MBD with GW in App.~\ref{sec:MBDGW}. We also investigated constraints from GW170817 \cite{Abbott:2018lct}, but they were weaker than the BBH catalog by about a factor of 5.

On the other hand, we plot results from our previous paper which examined testing massive scalar fields with SEP violation constraints \cite{Seymour:2019tir}. In particular, we studied the hierarchical triple PSR J0337 with an inner PSR-WD binary and a WD orbiting the outside (cyan). This configuration allows the constraints on SEP which we use to constrain MBD. 
Additionally, measurements of the consistency of the inverse square law (ISL) can be used to constrain MBD (dark purple). We plot these constraints from Ref.~\cite{Seymour:2019tir} for lunar laser ranging (LLR) \cite{Williams:2004qba} and planetary measurements \cite{Adelberger:2009zz}. 

In the end, the combination of ISL and SEP tests provide stronger constraints on MBD than binary PSR measurements. If a BH-PSR binary is found, it could provide similar constraints for low masses that PSR J0337 SEP constraints did. Additionally, a GW measurement of an asymmetric binary (i.e.~NS-BH) could provide strong constraints on MBD since the sensitivity difference would not be negligible like the NS-NS detection.

\section{Axion}\label{sec:axion}

\subsection{Theory}

The axion was first invented to solve the strong CP (charge-parity) problem \cite{Peccei:1977hh,Weinberg:1977ma,Wilczek:1977pj}. The theory of QCD does not preserve CP  symmetry generically, however experiments have found no evidence of CP violation in QCD. For example, measurements of the neutron's electric dipole moment failed to find CP violation \cite{Baker:2006ts}. The axion mechanism was introduced by promoting the QCD $\theta$ parameter to be a dynamical field with a potential so that it can naturally vanish and preserve CP symmetry. The axion is characterized by its mass $m_a$ and axion decay constant $f_a$. These parameters are related as
\begin{equation}\label{eq:QCDmasscouplingrel}
    m_a = 5.7 \times 10^{-12} \mathrm{eV} \left( \frac{10^{18} \mathrm{GeV}}{f_a} \right)\, ,
\end{equation}
to solve the strong CP problem \cite{diCortona:2015ldu}.

Moreover, there is significant physical motivation to consider a generalization of the QCD axion. These are referred to as axion-like-particles (ALPs) and do not have the mass $m_a$ and decay constant $f_a$ relationship in Eq.~\eqref{eq:QCDmasscouplingrel}. The compactification of string theories suggests the existence of ultralight ALP \cite{Arvanitaki:2009fg}. ALPs could also describe dynamical dark energy \cite{Kamionkowski:2014zda} and are a candidate for dark matter \cite{diCortona:2015ldu,Preskill:1982cy,Abbott:1982af,Dine:1982ah}. From now on, we will use axion to refer to both the QCD axion and ALPs. For these reasons, we are highly motivated to constrain axion parameter space with observations.

Axions show up in binary PSRs because stellar objects can acquire non-vanishing axion charges due to their sufficiently high matter densities. In contrast, axions are not sourced by e.g.~atomic nuclei \cite{Hook:2017psm}, so binary PSRs are a great laboratory for testing axions. When $m_a r \ll 1$, the axion field modifies the dynamics of the binary by creating a Yukawa potential between the two bodies of the form in Eq.~\eqref{eq:V_calG} with \cite{Hook:2017psm,Poddar:2019zoe}
\begin{equation}
 \mathcal{G} \equiv \left(1+ \frac{q_1 q_2 }{4\pi }\right) \, ,
\end{equation}
where $q_i$ is the dimensionless charge of body $i$ and has the form \cite{Poddar:2019zoe}
\begin{equation}\label{eq:AxionCharge}
    q_i = -\frac{8 \pi  f_a}{ \sqrt{\hbar}\ln{\left(1-\frac{2m_i }{R_i}\right)}} \, ,
\end{equation}
for mass $m_i$, and stellar radius $R_i$. Previous work \cite{Hook:2017psm,Huang:2018pbu} has given the axion charge as $q_i \sim 4 \pi f_a R_i / m_i$. This is related to Eq.~\eqref{eq:AxionCharge} by assuming a compact object ($m_i / R_i \ll 1$), then one can arrive at the previous definition. We will use the general expression for the axion in Eq.~\eqref{eq:AxionCharge} in our calculations. The axion charges are only non-vanishing for sufficiently high stellar densities, and the conditions are
\begin{align}\label{eq:axionchargecond}
    \rho_i \gtrsim m_a^2 f_a^2 \, , \quad
    \sqrt{\rho_i} \gtrsim \frac{f_a}{R_i} \, ,
\end{align}
otherwise, the axion charge is vanishing.

\subsection{Bounds}

Measurements of binary PSRs can be used to constrain the mass and coupling constant of an axion. Here, we will once again use the orbital decay rate observable to constrain deviation from GR without an axion presence. 
Thus, scalar dipole radiation will create a sizable deviation from standard GR in the orbital decay rate observable. Also, conservative modifications to the potential can occur as already explained, which will allow us to constrain some other regions.

We begin by calculating the modification of $\dot P$. The power radiated in a binary system consists of gravitational quadrupolar radiation and scalar dipolar radiation which are calculated to be \cite{Hook:2017psm} 
\begin{subequations}
    \begin{align}
        \dot E_Q &= - \frac{32}{5} \frac{\mathcal{G}^2 M^2 \mu^2 v^2 }{r^4} \, , \\
        \dot E_{sD} &= - \frac{1}{24 \pi}\frac{\mathcal{G}^2 M^2 \mu^2 \mathcal{S}^2}{r^4}\left(1-\frac{m_{a}^{2}}{\omega^{2}}\right)^{3 / 2} \Theta\left(\omega^{2}-m_{a}^{2}\right) \, ,
    \end{align}
\end{subequations}
where we define $\mathcal{S}^2 \equiv\left(q_1-q_2\right)^{2}$.
Using the relationship between energy lost and orbital decay rate in Eq.~\eqref{eq:pdot-generic} and the modified Kepler's law, we find the leading corrections to the orbital decay rate as in Eq.~\eqref{eq:mbdodrconstraint} with 
\begin{align}
  \kappa_1 \equiv& 12 \mathcal{G}^2 \, , \\
  \kappa_D \equiv& \frac{1}{8\pi}\mathcal{G}\left(1-\frac{m_a^2}{\omega^2}\right)^{3/2}\Theta(\omega-m_a) \, . 
\end{align}
This result will allow us to bound the axion with binary PSR observations.

\begin{figure}[htbp]
    \centering
    \includegraphics[width=8.5cm]{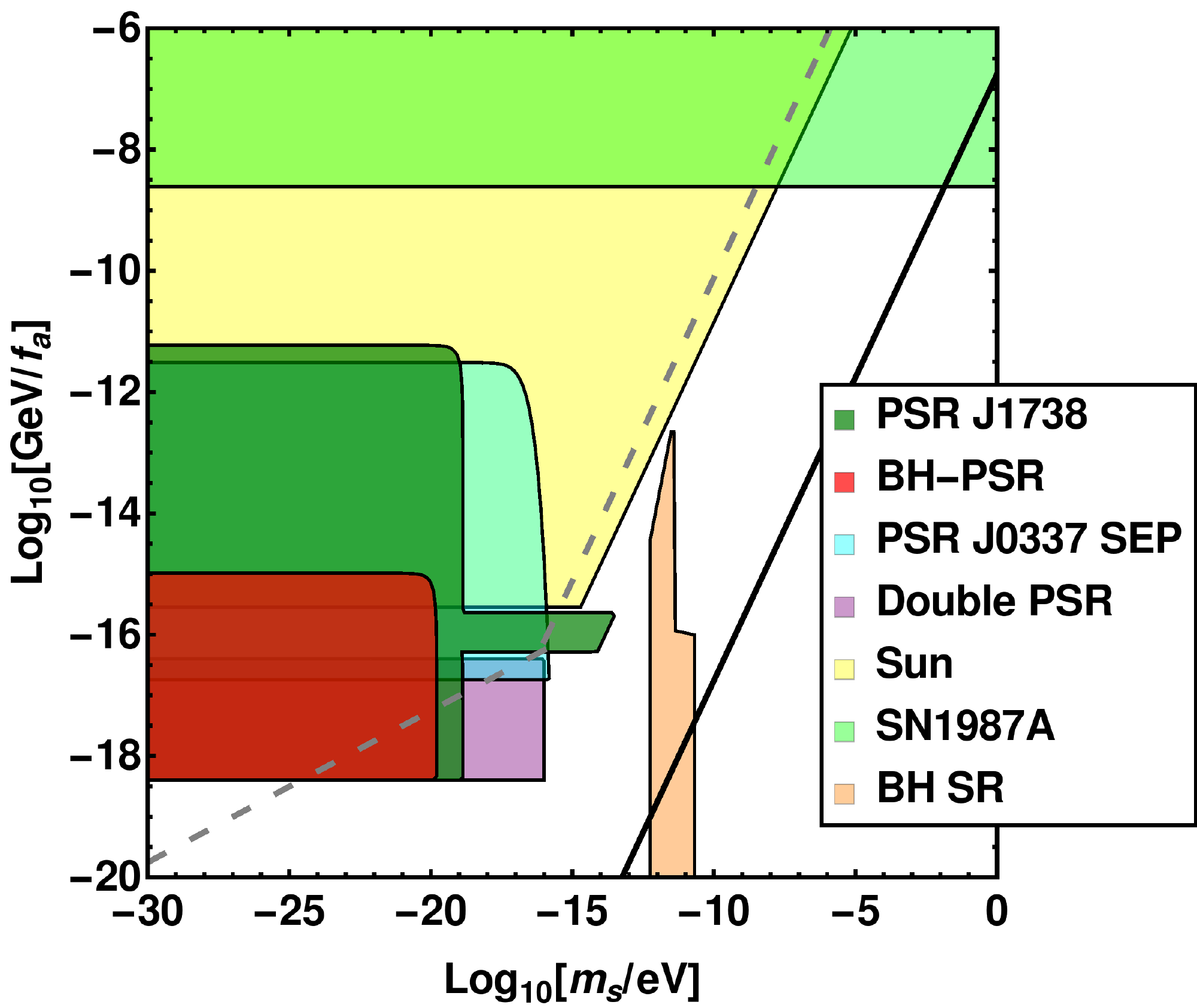}
    \caption{Constraints possible for the axion using PSR-WD J1738 and BH-PSR \cite{Freire:2012mg}. 
    We show the parameter space disallowed by PSR J1738 (dark green) and BH-PSR measurements with SKA (red).
    SEP measurements with the hierarchical PSR triple system PSR J0337 (cyan) also give constraints  \cite{Archibald:2018oxs, Seymour:2019tir} (subsequent analysis has further improved the J0337 SEP constraint by 30\% \cite{Voisin:2020lqi}). 
    We also show the constraints from measurements of the sun (yellow), supernova SN1987A (light green), and the absence of black hole superradiance (BH SR) (orange) \cite{Hook:2017psm,Arvanitaki:2014wva,Arvanitaki:2010sy}. 
    If the axion is the dark matter particle, then BBN measurements disallow the region above the dashed gray line \cite{Blum:2014vsa}.
    The black line represents the axion with finely tuned parameters that solve the strong CP problem (i.e. the axion is the QCD axion) \cite{Huang:2018pbu}. 
    The BH-PSR constraints are weaker than PSR J1738 because a BH has no scalar axion charge, so the system deviates less from GR. 
    }
    \label{fig:Axion}
\end{figure}

Using Eq.~\eqref{eq:mbdodrconstraint}, we construct bounds for the axion. In Fig.~\ref{fig:Axion}, we show the constraints possible using PSR J1738 (dark green) \cite{Freire:2012mg}. We also show the constraints with a BH-PSR binary (red) based on its simulated detection \cite{Liu:2014uka}\footnote{Others have shown that BH superradience can be used to constrain axions in a BH-PSR binary \cite{Kavic:2019cgk}. }. The constraints from these binary PSRs mainly come from those on dipolar radiation (which cuts off when $m_a = \omega$). The small sliver in J1738 comes from quadrupolar radiation modification, while the BH-PSR does not have this since the BH charge is zero. We also add previous constraints from solar \cite{Hook:2017psm}, supernova SN1987A, and BH superradiance constraints \cite{Arvanitaki:2014wva,Arvanitaki:2010sy}. We see that our constraints are largely similar to those of previous PSR constraints, i.e. the double PSR \cite{Hook:2017psm} and SEP violation constraints with PSR J0337 \cite{Seymour:2019tir}. We also show the constraints on the axion by Big Bang Nucleosynthesis (BBN) if it is the dark matter particle \cite{Blum:2014vsa}. Finally, we plot the relationship between the axion decay constant and axion mass from Eq.~\eqref{eq:QCDmasscouplingrel} if the axion is the QCD axion.

\section{Bound Dark Matter}\label{sec:bounddm}
 
\subsection{Theory}

With the introduction of dark matter, it is natural to ask whether dark matter can interact with itself. We refer to this interaction as the dark force. It is well theoretically motivated that dark matter can be gravitationally bound inside a NS or WD \cite{Goldman:1989nd,Kouvaris:2010vv,Bramante:2017xlb}. Thus, with the accumulation of excess bound dark matter in compact objects, a dark force between bodies in a PSR binary could be non-vanishing. In particular, we will consider a dark force that is described by a massive mediator.

Measurements of binary PSRs can thus be used to search and constrain the existence of a dark force. A dark force would change the orbital evolution of the binary and cause both conservative corrections from modification of the Newtonian potential and dissipative corrections from dipole and quadrupole radiation. While we are probing dark forces with PSRs, they have a strong precedent for many types of dark matter investigations. Binary PSRs near the galactic center are used to study the dark matter halo through dynamical friction \cite{Pani:2015qhr,Gomez:2017wsw}. Binary PSRs have also been used to probe oscillating dark matter distributions and to constrain a coupling between dark matter and standard model \cite{Blas:2019hxz,LopezNacir:2018epg}. On the other hand, there have been constraints on the dark force with binary PSRs for a particular dark matter model through modifications of the conservative sector (periastron advance) \cite{Fabbrichesi:2019ema}. GWs can also probe the dark force through both conservative and dissipative modifications from bound dark matter \cite{Ellis:2017jgp,Croon:2017zcu,Alexander:2018qzg}. 

We consider a generic way of constraining dark forces~\cite{Alexander:2018qzg}. This is because we are considering a widely separated binary in which the EFT of the models would cause similar phenomenology. However, we will make use of an example to be more explicit. We could have a model of asymmetric dark matter coupled to an Abelian gauge field \cite{Petraki:2013wwa,Petraki:2014uza,Kopp:2018jom}. Then a Lagrangian could be written as 
\begin{equation}
    \mathcal{L}_{\DM}=-\frac{1}{4} V_{\mu \nu} V^{\mu \nu}+\frac{1}{2} m_{v}^{2} V_{\mu} V^{\mu}+\bar{\chi}\left(i \gamma^{\mu} D_{\mu}-m_{\chi}\right) \chi
\end{equation}
with dark matter fermion $\chi$ with mass $m_\chi$, dark photon $V^\mu$ with mass $m_v$, gauge covariant derivative $D_\mu$, and dark photon field strength tensor $V_{\mu \nu}$.  Then, the dark photon will mediate a dark force between gravitationally bound dark matter inside NSs.

For a widely separated binary, we can consider the constituents as approximately point masses. The dark photon's interaction can be characterized as tree-level scattering, which allows us to describe it with a Yukawa potential in Eq.~\eqref{eq:V_calG} \cite{Alexander:2018qzg}. Once again, we will consider a very light particle so that orbital separation is much smaller the characteristic length of the dark force $m_v r \ll 1$. Then, binary's effective gravitational constant is given by
\begin{equation}
\mathcal{G} = 1-\alpha\,, \quad \alpha \equiv q_1 q_2\,,
\end{equation}
where $\alpha \equiv q_1 q_2$, $q_i$ is the dimensionless charge of body $i$. An antisymmetric dark matter model requires the interaction to be repulsive ($\alpha > 0$ ). We will see that we can constrain the maximum amount of dark matter charge in a typical $m\sim 1.4 \Msolar$ NS. This is what we refer to as a ``generic test" of dark matter since the relationship between dark charge and theory parameter depends on the dark matter model.

\subsection{Bounds}

Now, we derive the deviation of the GR orbital decay rate for a binary with a dark force. Using standard arguments, energy loss due to GW emission and vector dipole radiation are \cite{Alexander:2018qzg}
\begin{align} \label{eq:dmradiation}
    \dot E_{Q} &= - \frac{32}{5} \frac{\mathcal{G}^2 M^2 \mu^2 v^2}{r^4} \, , \\
    \dot E_{vD} &= - \frac{2}{3} \frac{\mathcal{G}^2 \mathcal{S}^2 \mu^2 M^2 }{r^4} \Theta\left(  \omega -m_v \right) \, ,
\end{align}
where $\mathcal{S}^2 \equiv \left( q_1 - q_2 \right)^2$ as in the axion case\footnote{The vector dipole radiation differs by a factor of two from previously examined scalar dipole radiation expressions \cite{Krause:1994ar,Alexander:2018qzg}.}. 
Using Eq.~\eqref{eq:pdot-generic} with the expressions for radiation with the dark force in Eq.~\eqref{eq:dmradiation}, the leading modifications of orbital decay rate are described by Eq.~\eqref{eq:mbdodrconstraint} with 
\begin{align}
    \kappa_1 &\equiv 12 \mathcal{G}^2 \, , \\
    \label{eq:dmkappas}
    \kappa_D &\equiv 2 \mathcal{G} \Theta\left(  \omega -m_v \right) \, .
\end{align}
From this, we see that the orbital decay rate is modified by both the change in total energy at 0PN order (and hence change in the quadrupolar radiation) and the vector dipole radiation term at $-1$PN order. 

With our expression for orbital decay rate modification with a dark force, let us now use binary PSR measurements to place constraints on it. We first specialize the orbital decay rate modification for BH-PSR and PSR-WD binaries. 
A BH has a vanishing dark charge because Bekenstein showed that a BH has vanishing charges for a scalar field or massive vector field \cite{Bekenstein:1971hc}. Additionally, the amount of bound dark matter in a NS will be much greater than that of a WD. This is a consequence of the dark matter capture rate being proportional to baryon density and escape velocity squared \cite{McDermott:2011jp}, and thus bound dark matter in NS dominates that of WD, $q_{\NS} \gg q_{\WD}$. The result of all of this is that we can use $\alpha \approx 0$ for a BH-PSR or PSR-WD binary. 

\begin{figure}[htbp]
    \centering
    \includegraphics[width=\linewidth]{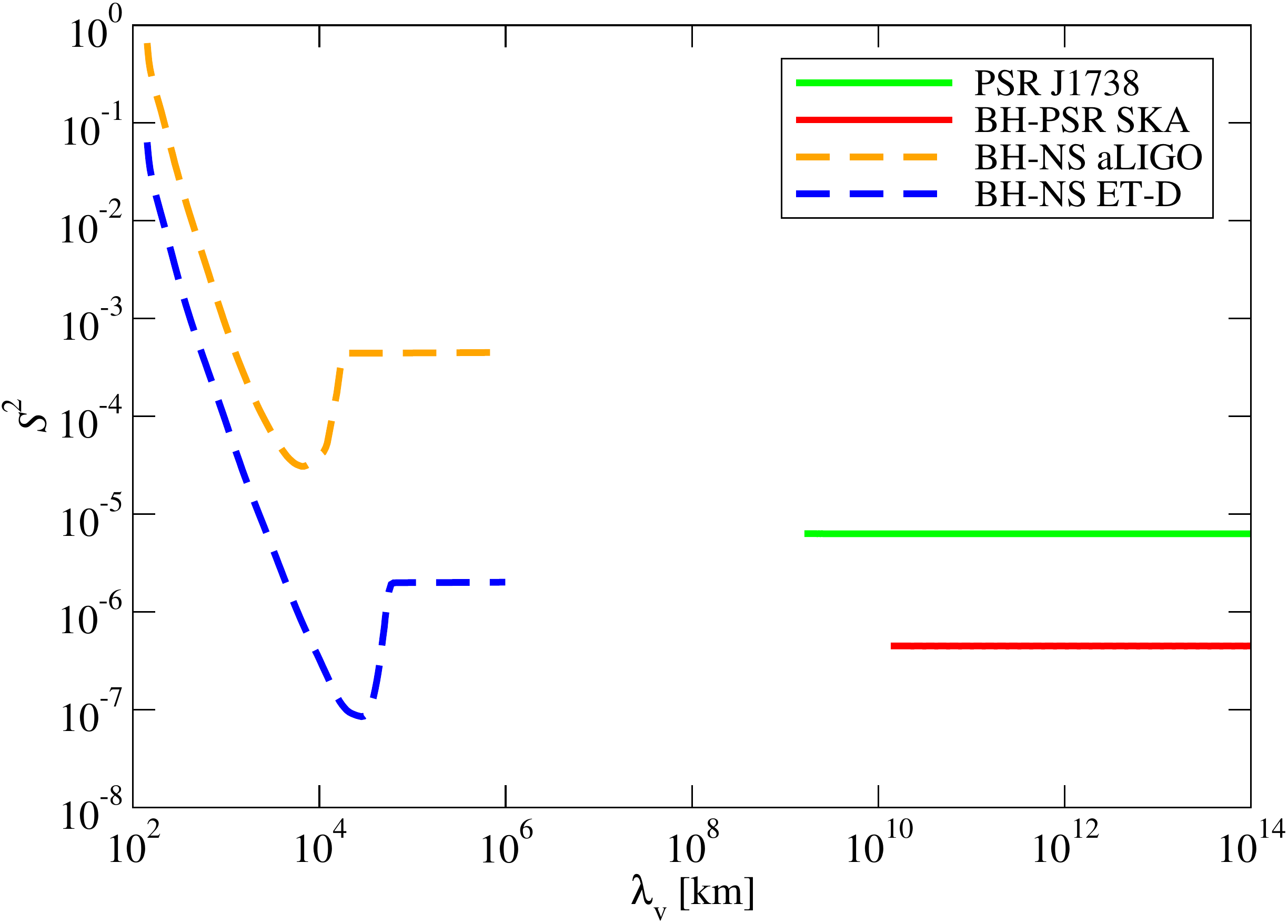}
    \caption{
        The upper bound on $\mathcal{S}^2$ as a function of dark force wavelength. We present the ones from PSR J1738 (green) and from the future detection of a BH-PSR binary with SKA (red). The cutoff of the PSR binaries occurs when the dipolar radiation ends ($\omega = m_v = \lambda_v^{-1}$ for the Compton length $\lambda_v$).
        The dashed lines give forecasted constraints on $\mathcal{S}^2$ for a future GW detection with aLIGO (dashed orange) and Einstein Telescope (dashed blue) \cite{Alexander:2018qzg}. The GW bounds use a binary of masses $(5,1.4)\Msolar$ at $150$ Mpc (full parameters used in Fisher analysis in Table~I of \cite{Alexander:2018qzg}). Furthermore, note that the GW bounds constrain up to $\lambda_v\sim 10^8$ km, but Ref.~\cite{Alexander:2018qzg} only shows up to $\lambda_v\sim 10^6$ km.
        One can see that measurement of binary PSRs is highly complementary to those made from GW.
        }
    \label{fig:dm-bounds}
\end{figure}

With these assumptions made, let us constrain dark forces with binary pulsar observations. With the expression for orbital decay rate modification in Eq.~\eqref{eq:mbdodrconstraint}, we can use measurements of PSR J1738 to place an upper bound on $\mathcal{S}^2 \approx q_{\NS}^2$. Similarly, we can use simulated measurement accuracies for a representative BH-PSR binary with SKA to place an upper bound on $\mathcal{S}^2 \approx q_{\NS}^2$\footnote{Note that the dark charge is dependent on the mass of the NS. However, we can show these in the same figure since both NS are approximately the same mass.}. Both of these constraints are shown in Fig.~\ref{fig:dm-bounds}, along with other constraints from GW observations as discussed in \cite{Alexander:2018qzg}. In both of these systems, the binary PSR constraints are complementary to those made with GW observations.  It can constrain the very light massive dark force part of the parameter space.

\section{Conclusion}\label{sec:conclusion}

In this paper, we have examined how one can use PSR binaries to constrain massive scalar/vector fields. We began by describing generic modifications of the orbital decay rate due to massive fields. We then used two representative systems to construct bounds: PSR-WD J1738 and a simulated BH-PSR. We found that for MBD, SEP and inverse square law constraints are stronger than binary PSR ones. For the axion, our constraints were similar to previous PSR-NS and SEP violation constraints for the theory parameter space. Finally, we saw that our binary PSR constraints on a massive dark force are complementary to previous GW results. 

In the future, binary pulsars will be of continued importance for testing massive fields. 
One important direction is a more rigorous analysis of pulsar timing data. Currently, we use parameter estimation results from the TEMPO package which assumes only GR. However, one could simultaneously estimate binary parameters and theory parameters to increase constraint accuracy (see e.g.~\cite{Anderson:2019eay}). 
More specifically, we could only probe the scalar dipole variable $\mathcal{S}^2$ in dark matter, but with more sophisticated analysis, one could constrain $\alpha$ by testing the consistency of advance rate of periastron, mass ratio, Shapiro time delay, and gravitational redshift with the double PSR J0737. This would likely improve on what is done in Ref.~\cite{Fabbrichesi:2019ema} which uses the Hulse-Taylor binary PSR B1913.
Recently, a new paper has derived the orbital decay rate in MBD without assuming a very light scalar mass ($m_s r \ll 1$) \cite{Liu:2020moh}. One could use their results to have a more accurate bound for more massive $m_s$. 

The continued discovery of new PSRs and instrumentation improvements mean that tests will become ever more precise. While we demonstrated how SKA will improve tests of massive fields with a BH-PSR binary, the next generation of radio telescopes could discover many types of exciting new pulsar systems. Pulsars will thus continue being part of the toolkit for probing fundamental physics.


\acknowledgments
 K.Y. acknowledges support from NSF Award PHY-1806776, NASA Grant 80NSSC20K0523, a Sloan Foundation Research Fellowship, and the Ed Owens Fund. 
K.Y. would like to also acknowledge support by the COST Action GWverse CA16104 and JSPS KAKENHI Grants No. JP17H06358.

\appendix

\section{MBD Gravitational Wave Constraints } \label{sec:MBDGW}

Let us now use GW observations to constrain MBD. Measurements of the gravitational waveform can be used to test the consistency of it with the GR waveform. To do this, we will use the parameterized post-Einsteinian framework~\cite{Yunes:2009ke} or the generalized IMRPhenom (gIMR) framework~\cite{TheLIGOScientific:2016src} which both generically describe phase and amplitude deviations to the waveform. The binary NS merger GW170817 has provided new constraints on deviation of the phase part of the gravitational waveform \cite{Abbott:2018lct}. Specifically, it has constrained the phase deviation at $0$PN and $-1$PN. We will see that the $0$PN measurement can constrain MBD most stringently because scalar dipole radiation at $-1$PN is negligible since both NSs have similar sensitivities.

First, let us describe the modification of the waveform phase in MBD. 
The Fourier transform of the gravitational waveform within the stationary phase approximation can be written as 
\begin{equation}
    \tilde{h}(f_\GW)= \mathcal{A} f_\GW^{-7 / 6} e^{\mathrm{i} \psi(f_\GW)} \, ,
\end{equation}
where $\mathcal{A}$ is an amplitude parameter, $f_\GW$ is the GW frequency, and $\psi(f_\GW)$ is the phase of the gravitational waveform \cite{Berti:2004bd}. This phase is computed in the stationary phase approximation (SPA) to be equal to \cite{Berti:2012bp}
\begin{widetext}
    \begin{equation}
        \psi(f_\GW) = 2\pi f_\GW t_c-\phi_c-\frac{\pi}{4}+\frac{3}{128 (\pi \mathcal{M}f_\GW)^{5/3}}\left\{1+\zeta  -\frac{5}{84}\eta^{2/5} (\pi \mathcal{M} f_\GW)^{-2/3} \xi \mathcal{S}^2 \Theta(\pi f_\GW -m_s)+ \mathcal{O}[(\pi \mathcal M f)^{2/3}]\right\} \, ,
    \end{equation}
\end{widetext}
where ($t_c$, $\phi_c$) are the time and phase at coalescence respectively, $\eta \equiv \mu/M$ is the symmetric mass ratio and $\mathcal{M}\equiv\eta^{3/5}M$ is the chirp mass\footnote{Note that the GW frequency is twice the orbital frequency $f_\GW=2f\propto v^3$, and thus $f_\GW$ is of $1.5$PN order.}.
$\zeta$ is defined as
\begin{equation}
    \zeta \equiv \frac{2}{3}\xi(s_1+s_2-2s_1s_2) +\frac{\xi}{2}-\frac{\xi \Gamma^2}{12}\Theta(2\pi f_\GW -m_s)\, .
\end{equation}

Since the GW frequency for the observed GW events is much higher than the orbital frequency of binary PSRs, we can assume $\pi f_\GW \gg m_s$ in our regime of interest. Namely, it is sufficient to consider the massless BD case. Then,
the modification of the waveform (up to 0PN order) reduces to
\begin{equation}
    \delta \psi_\MBD(f_\GW) \approx \zeta-\frac{5}{84}\eta^{2/5} (\pi \mathcal{M} f_\GW)^{-2/3} \xi \mathcal{S}^2 \, ,
\end{equation}
with
\begin{equation}
    \zeta \approx \frac{2}{3}\xi(s_1+s_2-2s_1s_2) +\frac{\xi}{2}-\frac{\xi \Gamma^2}{12}\, .
\end{equation}
These corrections to the GR waveform phase have been constrained from the observed GW events as
\begin{align}
    \left|\zeta\right| &< \delta \phi_0 \, , \label{eq:GWCons0PN}\\
    \Big| \frac{5}{84} \xi &\mathcal{S}^2 \Big| < \delta \phi_{-2} \,,
    \label{eq:GWCons-1PN}
\end{align}
where $\delta \phi_0$ and $\delta \phi_{-2}$ are the gIMR parameters at 0PN and $-1$PN orders respectively. With these inequalities relating MBD expressions with gIMR measurements, we can constrain MBD.

\begingroup
    \squeezetable
    \begin{table}
    \begin{centering}
        \begin{tabular}{|c|c|c|c|c| }
        \hline \hline 
        & \multirow{2}{*}{gIMR parameter  }& \multirow{2}{*}{MBD bounds  }     \\& &      \\ 
         \hline \hline
            \multirow{2}{*}{GW170817 \cite{Abbott:2018lct}  }& $\delta \phi_{0} < 3 \times 10^{-1}$& $\omega_\BD +3/2> 1.73$    \\ 
             & $ \delta \phi_{-2} < 2 \times 10^{-5}$ & $\omega_\BD +3/2> 0.64$   \\ 
            \hline
            \multirow{2}{*}{BBH  \cite{LIGOScientific:2019fpa} }& $\delta \phi_{0}< 9 \times 10^{-2}$& $\omega_\BD +3/2> 8.30$    \\ 
             & $\delta \phi_{-2}< 2 \times 10^{-3}$ & --    \\ 
        \hline
        \hline 
        \end{tabular} 
        \end{centering}
        \caption{\label{tab:GWgIMRbounds} A listing of GW constraints on MBD parameter. We list the gIMR parameter constraints from GW170817 and BBH cataloge GWTC-1. The MBD constraints are found from Eq.~\eqref{eq:GWCons0PN} and Eq.~\eqref{eq:GWCons-1PN} due to each gIMR parameter. 
        }
    \end{table}
\endgroup

The LIGO/Virgo Collaborations (LVC) has released papers putting constraints on gIMR parameters for both GW170817 \cite{Abbott:2018lct} and an analysis of BBH mergers \cite{LIGOScientific:2019fpa}. Starting with GW170817, we use the constraints on gIMR parameters $\delta \phi_0$ and $\delta \phi_{-2}$ as shown in Table~\ref{tab:GWgIMRbounds} \cite{Abbott:2018lct}. We use the low-spin priors and the median mass from GW170817 ($m_1 = 1.48 \Msolar$ and $m_2 = 1.265 \Msolar$). Reference~\cite{Wex:2014nva} calculated the sensitivities of NSs for typical equations of state, and we use the values of ($s_1 = 0.159$ and $s_2 = 0.140$) from Fig.~20. Considering constraints on both 0PN and $-1$PN with Eqs.~\eqref{eq:GWCons0PN} and~\eqref{eq:GWCons-1PN}, we calculate the bounds on MBD in Table~\ref{tab:GWgIMRbounds}. The strongest constraint is $\omega_\BD +3/2> 1.73$ at 0PN order because the difference in NS sensitivities is nearly vanishing for the scalar dipole radiation at $-1$PN. We next use the BBH merger analysis of gIMR constraints. For this, a BH sensitivity is $1/2$ so no scalar dipole radiation contributes. One can use the gIMR constraint in Table~\ref{tab:GWgIMRbounds} in conjunction with Eq.~\eqref{eq:GWCons0PN} to see that $\omega_\BD +3/2> 8.30$. Since BBH give the strongest bound from GW observations, we plot it in Fig.~\ref{fig:MBDOmegaBound}.

\section{Binary Parameters}\label{app:binary-param}

For completeness, we include all relevant binary parameters used in this paper's calculations in this section. They are listed in Table~\ref{tab:binary-values}. In this table, the binary masses are estimated by the PPK parameters assuming that GR is the correct theory. 
\begin{table*}
    \centering
    \begin{tabular}{|c|c|c|c|c|}
    \hline \hline
         &  J1738+0333&  J0348+0432 &  SDSS J0651+2844 &  SKA\\
    \hline \hline
    Binary Type & PSR-WD & PSR-WD & WD-WD & BH-PSR \\
    Reference & \cite{Freire:2012mg} & \cite{Antoniadis:2013pzd} & \cite{Hermes:2012us} & \cite{Liu:2014uka} \\
    \hline
        Period, $P$ (days) & 0.3548& 0.1024 & 0.008857& 3.0\\
        Orbital Decay Rate (intrinsic), $\dot P_\mathrm{intr}$  & $-25.9 \times 10^{-14}$& $-2.71 \times10^{-13}$ & $-8.2 \times10^{-12}$ &$-2.2\times10^{-14}$\\
        Primary Mass (PSR), $m_1$ ($\Msolar$)  & $1.46^{+0.06}_{-0.05}$& $2.01^{+0.04}_{-0.04}$ & $0.26^{+0.04}_{-0.04}$ &1.4\\ 
        Companion Mass (BH or WD), $m_2$ ($\Msolar$)  & $0.18^{+0.06}_{-0.05}$& $0.172^{+0.03}_{-0.03}$ & $0.50^{+0.04}_{-0.04}$ &10\\ 
        Fractional measurement error in $\dot P$, $\delta_\STAT$ & 0.13 & 0.18 & 0.34 & 0.019\\
        Fractional systematic error in $\dot P$, $\delta_\SYST$  & 0.057 & 0.046 & 0.20 & 0.0 \\
    \hline \hline
    \end{tabular}
    \caption{
    Parameters of the binaries considered in this paper. We show period, intrinsic orbital decay rate, masses, and both fractional measurement/systematic error in $\dot P$. Period is measured to at least nine digits, so we include the first four significant figures. For BH-PSR, we place the binary parameters according to the simulation in Ref.~\cite{Liu:2014uka}.
    }\label{tab:binary-values}
\end{table*}


\bibliography{bibliography.bib}

\end{document}